\documentstyle[12pt,aasms4]{article}

\begin{document}

\title{The Bull's-Eye Effect: Are Galaxy Walls Observationally Enhanced?}

\author{Elizabeth A. Praton\altaffilmark{1},
        Adrian L. Melott\altaffilmark{2}, and
        Margaret Q. McKee\altaffilmark{1}}

\altaffiltext{1}{Department of Physics, Grinnell College,
        Grinnell, IA  50112}
\altaffiltext{2}{Department of Physics and Astronomy, University of
        Kansas, Lawrence, KS  66045}

\begin{abstract}

We investigate a distortion in redshift-space which causes
galaxies to appear to lie in walls concentric about the observer, forming
a rough bull's-eye pattern.

We simulate what an observer would see in a thin slice of
redshift-space, including a magnitude limit
and constant slice angle.  The result is an enhanced ring of galaxies
encircling the observer at a distance roughly corresponding
to the peak of the selection function.
This ring is an artificial enhancement of weak
features in real-space.
This may explain visually prominent features such as the ``Great
Wall'' and periodicity found in deep narrow fields.

\end{abstract}

\section{Introduction}

Most maps of large scale structure use
redshifts of galaxies to indicate position.  Such maps are known to have
distortions, the most familiar of which are the so-called
fingers-of-god, elongated artifacts pointing at the observer and
caused by velocities of galaxies within clusters. On scales larger than
a typical finger-of-god, the maps show other features, such as great
connected walls of galaxies.  On these scales, are redshift-space maps
essentially accurate representations of the real-space distribution?
Can they be treated as fuzzy photographs of the universe?

In this paper we show that the answer
may be {\em no}. A spectacular feature
seen in redshift surveys---the curving ``Great Wall'' of galaxies which
together with the Pisces-Perseus chain seems to
form a giant ring encircling us
(\cite{GH89}; \cite{dC94}; \cite{MHG96})---may get its visual
punch from a redshift-space distortion we call the bull's-eye effect.

Using an N-body simulation, we find that structures
perpendicular to the line of sight are enhanced in redshift-space,
due both to the finger-of-god distortion and a distortion caused by infall.
We can further strengthen the effect by applying a magnitude limit.
The result is that any observer concludes, erroneously, that
he or she is encircled by galaxy walls in a bull's-eye pattern.

\section{Redshift-Space Artifacts}  \label{sec:artifacts}

When peculiar velocities are small and uncorrelated, redshift maps are just
fuzzier versions of real-space maps.  However, large correlated
velocities produce striking artifacts.  One example of such an artifact
is the finger-of-god produced by the random orbital motions of galaxies
in a bound cluster.   Another artifact, not as well known but important
for this paper, is caused by the inward flow of material toward either a
single center or an extended region.

Such a flow produces an artifact because infalling material on the near-side
of the accreting center or extended region has peculiar velocity away from the
observer and material on the far-side has peculiar velocity toward the
observer.  Therefore, in redshift-space, the near-side is moved back
and the far-side moved forward, squashing the material together along
the radial direction.  For example, studies show that spherical
infall produces an extended saucer-like artifact face
on to the observer, encircling the cluster finger-of-god like an
outspread skirt or tutu (\cite{K87}; \cite{RG89}; \cite{PS94}, hereafter
PS).  These results are generalizable
to aspherical flows.  It is easy to show that
{\em any} infall field will produce a similar artifact:  see the discussion
of Fig.~1 in PS.

Kaiser (1987) speculates that infall distortions may be ``well able to account
for'' connected features seen in redshift surveys.  Likewise, the
schematic Fig.~6 of PS shows how an observer in
a universe with many clusters might see a bull's-eye pattern in
redshift-space. More realistic scenarios than the isolated cluster model
can produce an even stronger effect:  bull's-eyes are present (but
uncommented on)
in figures presented by Park(1990) and Ryden \& Melott (1996; hereafter, RM).
Weinberg \& Gunn (1990a;b) comment on ``hints of circular symmetry''
visible in their plots (which show the effect less clearly), but attribute
it to the selection function.  As we will show below, the full bull's-eye
effect is a true distortion in redshift-space, and the selection function
plays only a secondary role.

We wish to emphasize that walls and filaments are {\em real} features
that arise from long-wave motion producing ``pancakes'' by gravitational
instability (\cite{MS90}; \cite{PM95}) and so do not
require additional explanation
such as in Kaiser (1987).  On the other hand, we find that redshift effects
selectively enhance pancakes forming perpendicular to the line of sight,
accelerating their apparent collapse.
The toy model of an oblate spheroid may be useful here. Collapsing with its
short axis along the line of sight, it may look very thin and dense in redshift
space. With short axis perpendicular to the line of sight, redshift distortions
make it look smaller and more nearly spherical.

\section{A Simulated Universe}
\label{sec:simulation}

Before comparing redshift surveys with
simulations of gravitational clustering, the two sets of
data should be transformed to the same coordinate system.
Observers usually work with selection effects in redshift-space, whereas
simulations give complete information in real-space. Since it is
presently impossible to
deconvolve an observed data set to a complete real-space distribution,
we must simulate realistic observational constraints, as done by Park (1990)
and Weinberg \& Gunn (1990 a;b).

The simulation we choose to look at is from a set of two-dimensional
studies described in detail
in Beacom et al. (1991) and Kauffmann \& Melott (1992).
RM calculated the properties of voids in
three simulations in real-space and redshift-space.  Of the three, the one with
initial power spectrum of the form $P(k) \propto k^0$ shows the
bull's-eye effect most strongly.
This index corresponds to $P(k) \propto k^{-1}$ in three dimensions, which
is a reasonable approximation to the observed power spectrum over a
wide range of scales (\cite{GR75}; \cite{FKP94}; \cite{PD94};
\cite{LKSLOTS96}).
 For these spectra, displacement varies logarithmically
with $k$ (\cite{S93}), which can cause structure to ``pile up''.
The main effect of the reduced dimensionality of
the simulation is the absence
of the occasional very long finger-of-god due to clusters like
Coma seen in the first CfA slice (\cite{dLGH86}).

Figure~\ref{fig:twoobs}(a) shows this simulation in real-space,
and Fig.~\ref{fig:twoobs}(b) in redshift-space,
as seen by two hypothetical observers whose positions are
labeled on the plot.  The simulation has periodic
boundary conditions, so the figure shows the original points tiled
together with copies.  As in RM, we've first reduced the number
of particles by randomly selecting one out of every four.
The square has a side of approximately $600 h^{-1}$ Mpc, where $h$ is the fudge
factor in the Hubble constant:
$H_0 = 100 h \mbox{ km }\mbox{s}^{-1}\mbox{ Mpc}^{-1}$, and where
we let $h = 1$ for convenience. The assumed value of $h$ affects only
the selection function, not redshift displacement.
The redshift-space maps made by
the observers each have a radius of about $30,000 \mbox{ km~s}^{-1}$
and overlap as indicated.  Each
map includes a selection function simulating
the effect of a magnitude 16 limited survey.

Specifically, the luminosities of the points are assumed to obey
a Schechter luminosity function:
$N(>L/L_*) \propto \Gamma(1+\alpha, L/L_*)$, where we use
$\alpha = -1.07$ and $L_* = 1.0 \times 10^{10} h^{-2}\mbox{~L}_{\sun}$
(e.g., Peebles 1993).
We also assume that the ``volume'' of the observer's sample increases
with distance as in a constant angle slice.  That is, we randomly select
points so that the surface density at distance $r$ goes as
$\sigma \propto r \, \Gamma(1 + \alpha, L_0(r)/L_*)$, where $L_0(r)$ is the
luminosity corresponding to the magnitude limit at the distance $r$.

Note that when we apply the selection function above, we
essentially treat the two-dimensional simulation as if it
were a three-dimensional slab.  This is admittedly a swindle, but
we believe a not inaccurate approximation for a thin slice.
Our maps are similar in
appearance to those in Park (1990), which were constructed from a
three-dimensional model.
Due to the high resolution of our simulation,
the number of points exceeds the number of galaxies that would lie in
a thin slice. Since we wish to show the origin of the effect as clearly
as possible (rather than compare our results with specific surveys),
we have normalized the selection function
to keep all the subset simulation particles in the region of its maximum.

The difference between the real-space distribution in
Fig.~\ref{fig:twoobs}(a) and the redshift-space maps in
Fig.~\ref{fig:twoobs}(b)
is readily apparent.  In the first, the distribution of ``galaxies''
is fairly uniform, whereas in the second, the galaxies seem to lie in
curving walls which roughly circle the observer, regardless
of position!  Why are the two maps so different?

Figure~\ref{fig:unselsel} shows the simulation from the point of view
of the first observer in more detail.  Note:
(1)  The positions of the galaxies are distorted in redshift-space into
smeared arclets concentric around the observer.  Figure~\ref{fig:unselsel}(a)
shows the points plotted in real-space and redshift-space without the
selection function. This pair of plots should be compared with Fig.~2 of
RM, which shows a similar pattern appears for an observer
at the center of the original square.  (2) The selection effect of the
magnitude limit enhances arclets at the
distance where the selection function peaks.  Figure~\ref{fig:unselsel}(b)
shows the real-space and redshift-space distribution of the points when
the selection function is applied.  This time the magnitude limit is
17, to show the effect of an increase
(compare with Fig.~\ref{fig:twoobs}).  The selection function
is picking out bull's-eye rings already existing in redshift-space:
it does not {\em produce} the pattern, as suggested by
Weinberg \& Gunn (1990a;b).

\section{Underlying Mechanism}
\label{sec:mechanism}

The mechanism underlying the bull's-eye effect is redshift-space
distortion caused by peculiar velocity.
Both infall onto clusters and orbital motions within
clusters contribute, as can be seen in an expanded piece of the simulation.
Figure~\ref{fig:wedge} shows a $20\arcdeg$ wedge of the simulation in
real-space and redshift-space, from the point of view of an observer
at the center of the original square (Fig.~2 of RM) looking north.  It shows
several points clearly:

1. The distortion due to infall tends to empty voids and pancake galaxies
on top of each other in redshift-space, perpendicular to the line of sight.
Note how the two filaments on the left-hand side of the wedge
between 100 and 150 Mpc are squashed together in redshift
space.

2.  The finger-of-god artifacts give structures lying perpendicular
to the line of sight more visual weight, by smearing the points out along the
line of sight.  A careful inspection of the loose filament running
along the top of the wedge between
250 and 300~Mpc reveals that points are
just smeared out from their positions right on top of each other.
Furthermore, the fingers-of-god can be seen to be longer in dense regions
with more tightly bound clumps.  This correlation further increases the
visual contrast.

3. The bull's-eye pattern appears because only structures perpendicular to
the line of sight are enhanced.  The long filament running down the center
of the
wedge from 200 to 300~Mpc is not particularly enhanced in redshift-space.
However, if the observer was located to the left or right of this
filament, it would appear darker and thicker than it does in this view.

Again, we reiterate that the effect is caused by both the infall compression
distortion and the more subtle finger of god smearing distortion. We have
looked
at two simulations in which one or the other distortion is suppressed.
In one, small scale clustering (and thus fingers of god) are suppressed:  the
result is a bull's-eye pattern which is not quite as striking as the ones shown
here, rather like the figures in Weinberg \& Gunn (1990a;b).  In the other,
peculiar velocities were reassigned randomly, removing infall:  the result
is a much more uniform pattern with only extremely weak suggestions of walls.

Note that the velocities in our simulation are not large.
A few hundred km/s displacement is enough to produce the effect.

\section{Conclusions}
\label{sec:conclusions}

In the simulation examined above
both motions within clusters and flows toward clusters
produce distortions in redshift-space which enhance structures
perpendicular to the line of sight.  Redshift maps thus give the impression
that the observer is surrounded by concentric walls of galaxies:
the bull's-eye effect.

It is not implausible that the bull's-eye effect may be responsible for
the visual impact of wall-like features concentric
about our position, such as the Great Wall of galaxies.
We would not argue that there is no physical structure
underlying these features, but that existing structure may be enhanced
by a simultaneous smearing and compressing distortion as in the
simulation.

We note that a full view of the simulation in redshift-space
shows multiple rings, separated by fairly uniform
intervals.  A corresponding prediction is that as the magnitude limit of
magnitude complete redshift surveys is increased, other sets of walls
concentric on our position but at larger distances should become apparent.
Indeed, the map produced by the Las Campanas survey (\cite{LSLKOT96})
is similar in
appearance to Fig.~\ref{fig:unselsel}(a) above.
We also suggest a possible relationship to other findings
based on ``core-sampling'' (\cite{BEKS90}; \cite{DTOKLSLF96}; \cite{CHPB96}).

It is possible that the bull's-eye effect is giving a misleading impression
about the nature of the large scale structure of the universe.  The
distribution of galaxy clusters and filaments may be more uniform
than impressions from redshift-space maps imply.

We wish to emphasize that our simulation only demonstrates the effect
qualitatively.  A detailed comparison with observational data demands
the use of a three-dimensional simulation with a power spectrum more
closely matched to that observed in the local universe.
The spacing and intensity of this effect will depend on details of the power
spectrum and mass density of the Universe.
In fact, characteristic ring spacing combined with the
measured angular power spectrum should measure bulk flows
and provide information on $\Omega_0$.  Future work should include
examination of the effect with a variety of three
dimensional spectra and background cosmologies.

\acknowledgements

We wish to thank S. Schneider, I. Praton, S. Shandarin, D. Weinberg, and
J. Gott for helpful comments.  ALM wishes to acknowledge financial support
from NASA Grant NAGW 3832, the NSF EPSCoR program, and Space Telescope
Science Institute award number 6007.

\clearpage
%
%%%%%%%%%%%%%%%%%%%%
% Figure 1
%%%%%%%%%%%%%%%%%%%%
\begin{figure}
\caption{
(a) A map showing the true positions of all ``galaxies'' and
two observers in a two-dimensional
simulated universe with periodic boundary conditions.
The figure consists of four copies of the original set of points
(inset square) tiled together.
(b) The redshift-space maps made by the two observers.  These maps show
the velocity positions of all galaxies lying in a simulated slice with
a magnitude greater than 16.
\label{fig:twoobs}}
\end{figure}
%
%%%%%%%%%%%%%%%%%%%%%
% Figure 2
%%%%%%%%%%%%%%%%%%%%%
\begin{figure}
\caption{Maps of the simulation made by the first observer from
Fig.~\protect\ref{fig:twoobs}.  The observer is at the center of the circles.
(a) The simulation in real-space and red-shift space,
without any selection effect.
(b) The simulation as seen by a survey with a magnitude limit of 17, in
real-space and redshift-space.
\label{fig:unselsel}}
\end{figure}
%
%%%%%%%%%%%%%%%%%%%%%
% Figure 3
%%%%%%%%%%%%%%%%%%%%%
\begin{figure}
\caption{A wedge of the simulated universe in real-space and
redshift-space, as seen by an observer at the center of the inset
square in Fig.~\protect\ref{fig:twoobs}(a).
\label{fig:wedge}}
\end{figure}


\begin{thebibliography}{}

\bibitem[Beacom et al. 1991]{BDMPS91}
Beacom, J. F., Dominik, K. G., Melott, A. L., Perkins, S. P., \&
Shandarin, S. F. 1991, ApJ, 372, 351

\bibitem[Broadhurst et al. 1990]{BEKS90}
Broadhurst, T. J., Ellis, R. S., Koo, D. C., \& Szalay, A. S.  1990,
Nature, 343, 726

\bibitem[Cohen et al. 1996]{CHPB96}
Cohen, J. L., Hogg, D. W., Pahre, M. A., \& Blandford, R. 1996,
ApJ 462, L9

\bibitem[da Costa et al. 1994]{dC94}
da Costa, L. N. et al. 1994, ApJ, 424, L1

\bibitem[de Lapparent, Geller, \& Huchra 1986]{dLGH86}
de Lapparent, V., Geller, M. J., \& Huchra, J. P. 1986,
ApJ, 302, L1

\bibitem[Doroshkevich et al. 1996]{DTOKLSLF96}
Doroshkevich, A. G., Tucker, D. L., Oemler, A., Kirshner, R. P.,
Lin, H., Shechtman, S. A., Landy, S. D., \& Fong, R. 1996,  MNRAS in press

\bibitem[Feldman, Kaiser, \& Peacock 1994]{FKP94}
Feldman, H. A., Kaiser, N., and Peacock,
 J. A.  1994, ApJ, 426, 23

\bibitem[Geller \& Huchra 1989]{GH89}
Geller, M. J., \& Huchra, J. P. 1989, Science, 246, 897

\bibitem[Gott \& Rees 1975]{GR75}
Gott, J. R., \& Rees, M. J. 1975, A\&A, 45, 365

\bibitem[Kaiser 1987]{K87}
Kaiser, N. 1987, MNRAS, 227, 1

\bibitem[Kauffmann \& Melott 1992]{KM92}
Kauffmann, G., \& Melott, A. L. 1992, ApJ, 393, 415

\bibitem[Landy et al. 1996]{LSLKOT96}
Landy, S. D., Shectman, S. A., Lin, H., Kirshner, R. P., Oemler, A. A.,
\& Tucker, D. 1996, ApJ, 456, L1

\bibitem[Lin et al. 1996]{LKSLOTS96}
Lin, H., Kirshner, R. P., Shectman, S. A., Landy, S. D., Oemler, A.,
Tucker, D. L., \& Schechter, P. L. 1996, ApJ, 471, 617

\bibitem[Marzke, Huchra, \& Geller 1996]{MHG96}
Marzke, R. O., Huchra, J. P., \& Geller, M. J. 1996, AJ,
112, 1803

\bibitem[Melott \& Shandarin 1990]{MS90}
Melott, A. L., \& Shandarin, S. F. 1990, Nature, 346, 633

\bibitem[Park 1990]{P90}
Park, C. 1990, MNRAS, 242, 59p

\bibitem[Pauls \& Melott 1995]{PM95}
Pauls, J. L. \& Melott, A. L. 1995, MNRAS, 274, 99

\bibitem[Peacock \& Dodds 1994]{PD94}
Peacock, J. A., \& Dodds, S. J. 1994, MNRAS, 267, 1020

\bibitem[Peebles 1993]{P93}
Peebles, P. J. E. 1993, Principles of Physical Cosmology
(Princeton, NJ: Princeton University Press)

\bibitem[Praton \& Schneider 1994]{PS94}
Praton, E. A., \& Schneider, S. E. 1994, ApJ, 422, 46 [PS]

\bibitem[Reg\H{o}s \& Geller 1989]{RG89}
Reg\H{o}s, E., \& Geller, M. J. 1989, AJ, 98, 755

\bibitem[Ryden \& Melott 1996]{RM96}
Ryden, B. S. \& Melott, A. L. 1996, ApJ, 470, 160 [RM]

\bibitem[Shandarin 1993]{S93}
Shandarin, S. F. 1993, Proc. Cosmic Velocity Fields,
eds. F.R.Bouchet, M.Lachieze-Rey, Editions Frontieres, 383

\bibitem[Weinberg \& Gunn 1990a]{WG90a}
Weinberg, D. H., \& Gunn, J. E. 1990a, ApJ, 352, L25

\bibitem[Weinberg \& Gunn 1990b]{WG90b}
Weinberg, D. H., \& Gunn, J. E. 1990b, MNRAS, 247, 260

\end{thebibliography}
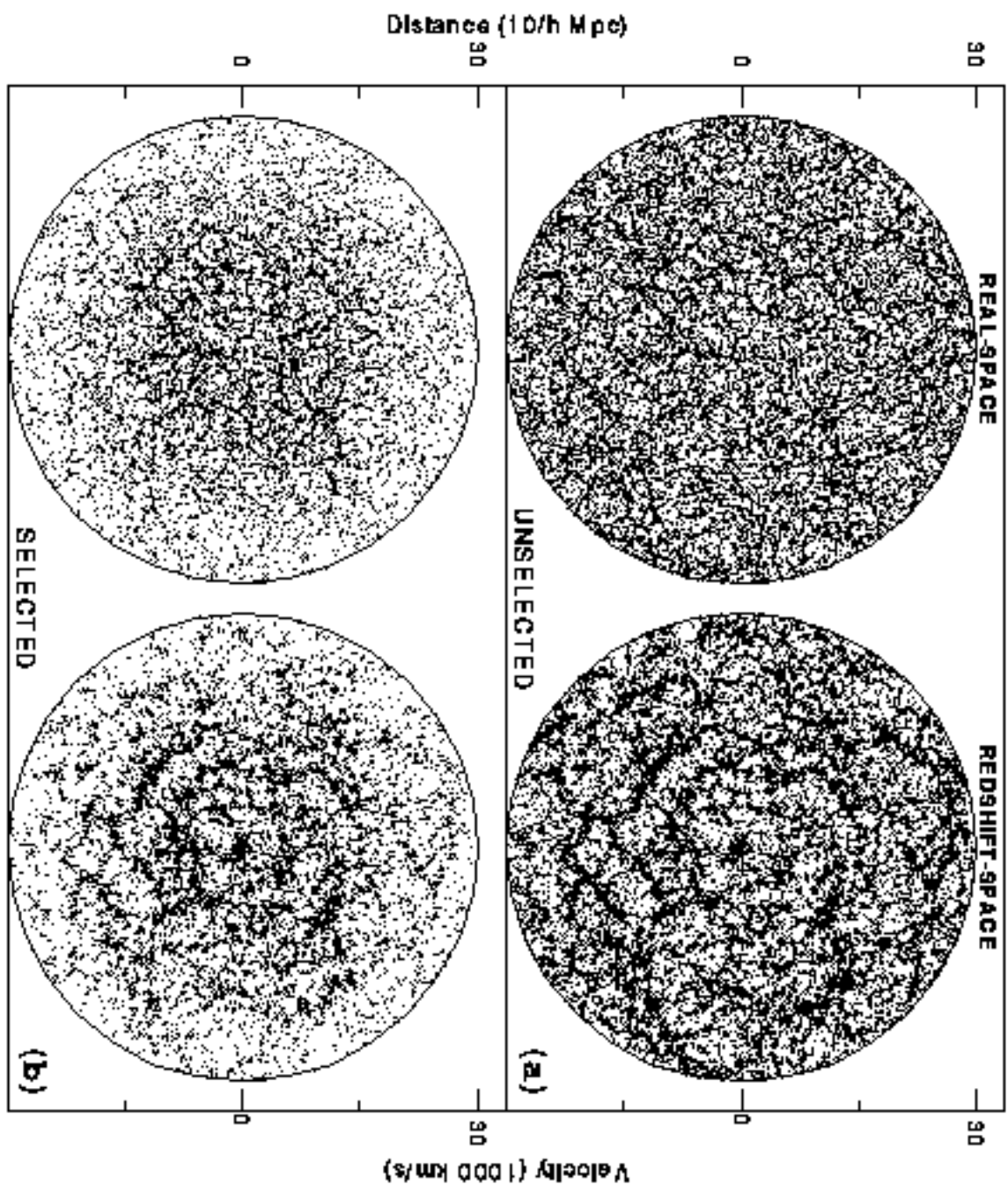
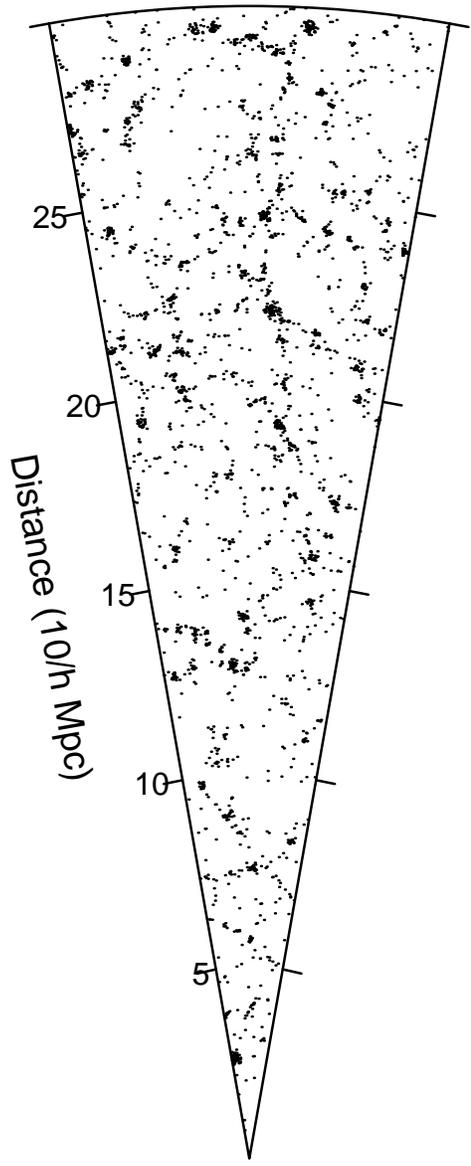
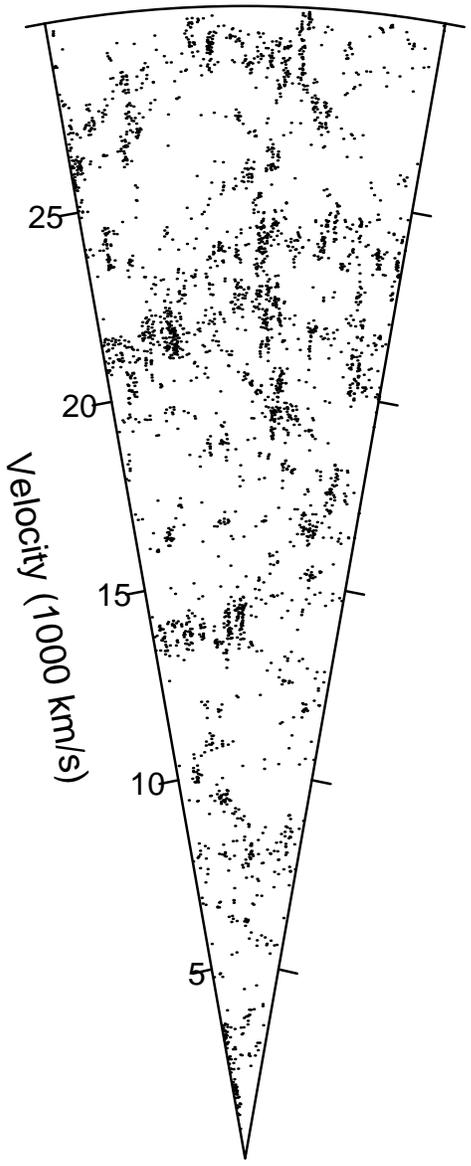
\end{document}